# External busbars for improving current generation in multijunction solar cells

Marianna Vuorinen[*], Arto Aho, Elina Anttola, Antti Fihlman, Ville Polojärvi, Timo Aho, Riku Isoaho, Veikka Nikander, Arttu Hietalahti, Antti Tukiainen, Mircea Guina

*Optoelectronics Research Centre, Physics Unit, Tampere University, P.O. Box 692, FI-33014 Tampere, Finland*
[*]*Corresponding author.*
*E-mail addresses: marianna.vuorinen@tuni.fi (M. Vuorinen), arto.aho@winsepower.com (A. Aho), elina.anttola@vexlum.com (E. Anttola), antti.fihlman@winsepower.com (A. Fihlman), ville.polojarvi@winsepower.com (V. Polojärvi), timo2aho@gmail.com (T. Aho), riku.isoaho@tuni.fi (R. Isoaho), veikka.nikander@winsepower.com (V. Nikander), arttu.hietalahti@vexlum.com (A. Hietalahti), antti.tukiainen@tuni.fi (A. Tukiainen), mircea.guina@tuni.fi (M. Guina)*

**Abstract**

We report an improved device fabrication process employed in the development of an advanced front contact grid design employing external busbars. The advanced fabrication process results in enhanced solar cell performance measured at one-sun illumination. In this grid configuration the busbar area is located outside the active solar cell and the grid fingers travel across the mesa sidewalls. With this design the solar cell size can be scaled down without limitations as the area of the busbar isn't restricting the component size. Thus, the demonstrated design is beneficial especially for micro-concentrator solar cells. In general, this approach minimizes the power losses originating from the grid shadowing and dark area related voltage losses. The performance of the proposed design is validated by fabricating GaInP/GaAs/GaInNAsSb triple-junction solar cells employing the grid design with external busbars. Light-biased current-voltage and electroluminescence characteristics of the cells reveal that an additional contact GaAs etching step prior to front contact metal deposition is needed to ensure good photovoltaic performance with a fill factor of 85% at one-sun illumination. The improvement is attributed to removing the plasma-damaged material layer that can extend to a depth beyond 100 nm, leading to resistive losses.

## 1. Introduction

III–V multijunction solar cells (MJSCs) are dominating the landscape for the highest photovoltaic (PV) performance among all PV technologies. To date, the highest conversion efficiency has been achieved with a four-junction III–V solar cell, reaching 47.6% at 665 suns concentration [1]. Key for achieving high-efficiency operation in terrestrial applications is deployment of concentrated photovoltaics (CPV), where the incident light is concentrated on a solar cell with the help of optics [2, 3]. Among many specific optimization aspects, this sets strict requirements also for the front contact grid design. Firstly, for high-efficiency CPV performance, electrical contacts need to meet the requirements of conducting high current densities. This can be ensured by fabricating grids with sufficiently high cross-sectional areas. At the same time, the losses arising from the shadowing caused by the front contact as well as all the resistive losses become more critical compared to non-concentrated solar illumination [4]. In fact, the contact grid pattern has a major effect since on one hand the losses originating from the shadowing are directly proportional to the grid area [4], and on the other hand minimizing the resistive losses would require a sufficiently dense grid.

A conventional, linear contact grid design for CPV solar cells includes grid fingers which collect charge carriers from the emitter and conduct them forward to busbars from where the current is directed to the external circuit. Since busbars collect the current from multiple fingers and act also as a contact area for the interconnects, their dimensions are typically larger than that of the grid fingers, increasing the shadowing significantly. In addition to collection loss resulting directly from the shadowing effect, the conventional busbars lead to power loss resulting from the reduction in the open-circuit voltage caused by the increase in the dark saturation current relative to the light-generated current, which also diminishes some of the efficiency gained with concentration [5]. At the same time, since the busbars contribute most to the overall contact grid area, the probability of the grid metal coinciding with an epitaxial defect, having detrimental effect on the device performance, is increased significantly with the conventional busbar design. To mitigate this, the busbars can be partially [6] or even completely [7–9] removed from the active area of the solar cell. In this work we present an advanced front contact grid design in which the busbars are located outside the active area, and only the fingers cause shadowing losses. Moreover, completely moving the busbar from the active solar cell area enables fabrication of smaller solar cells since dimensional limitations of busbars are not constraints for the cell size, making this especially beneficial for micro-scale CPV applications [10–12]. Furthermore, as a future prospect of the proposed grid concept, once the busbar area is successfully fabricated outside the mesa structure, the device configuration can be further developed towards an architecture in which the busbar area is fabricated completely outside the III–V material. Ultimately, this would translate into the reduction in the III–V material usage and costs.

We have previously demonstrated a proof-of-concept contact grid design for III–V MJSCs with the busbar placed outside the active region [13]. Here we report on the improvements in the fabrication process, resulting in significantly enhanced device performance. The operation of the device components was characterized by electroluminescence (EL) and current-voltage

(IV) measurements. In addition, contact resistance measurements were used for analyzing the electrical properties of the metal-semiconductor interface.

## 2. Experimental
### 2.1 Contact grid design and fabrication details

The schematic presentation of the grid design and placement of the busbar is shown in Fig. 1. Only the grid fingers are situated on the front surface of the solar cell as well as on the mesa facets, collecting the charge carriers from the emitter and further conducting the current to the busbars across the mesa facets. In order to electrically isolate the fingers from the solar cell facet, an insulating [13] dielectric double-layer coating is implemented on the mesa sidewall underneath the fingers. Due to fabricational limitations, a small section of dielectrics covers the edges of the front surface next to the mesa facets. This is referred to as a dielectric offset.

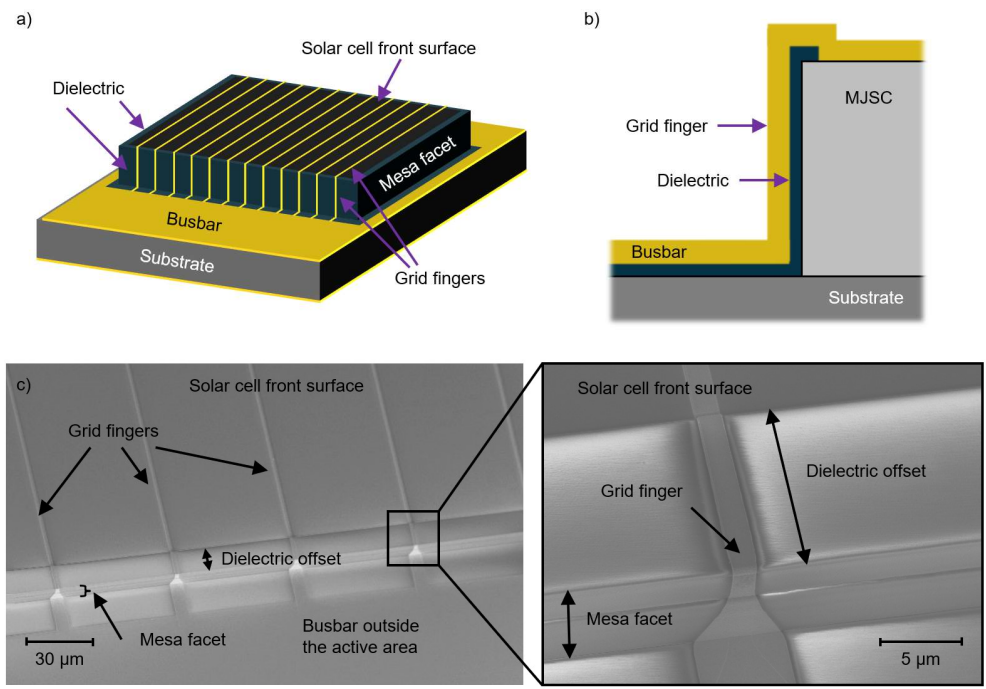

*Figure 1. a) Schematic of the device architecture with external busbars. b) Schematic showing the layer structure of the design. c) Scanning electron microscopy images of a final component. This particular component has a dielectric offset of 10 µm.*

Test lattice-matched monolithic solar cell structures were grown on 4" *p*-GaAs substrates with a Veeco GEN20 molecular beam epitaxy system. A more detailed description of the growth of dilute nitride solar cells can be found elsewhere [14]. The sample set consisted of two different GaInP/GaAs/GaInNAsSb triple-junction structures, denoted as SC1 and SC2. The main difference in the epitaxial structures is in the GaInP junction thicknesses, namely for SC2 the GaInP junction is thinner by 38% compared to SC1. Mesa structures of 2×2 mm$^2$ were fabricated by inductively coupled plasma etching. A dielectric double-layer $SiO_2/SiN_x$ (300 nm / 300 nm) was deposited on the entire front surface by plasma-enhanced chemical vapor deposition (PECVD). Photolithographic masking and reactive ion etching (RIE) with Ar and

CHF$_3$ as processing gases were used for etching the dielectrics from the front surface of solar cells, while leaving the dielectric double-layer on the mesa facets. Components with dielectric offsets ranging from 5 to 30 µm were fabricated to investigate how accurate alignment can be achieved with the mask aligner tool in use when fabricating the opening in the dielectrics. The front metal contacts (non-alloyed Ni/Au) together with back metal contacts (non-alloyed Ti/Pt/Au) and a double-layer TiO$_x$/SiO$_y$ antireflection coating (ARC) were deposited by electron beam evaporation. The front contact metallization was partially deposited in normal incidence and partially by tilting the samples 45° clockwise and counterclockwise during the deposition to obtain continuous metal coverage for the fingers across the mesa facets and the edges of the dielectric layers. The width of the contact grid fingers was approximately 2 µm. Before the ARC deposition the contact GaAs layer in between the fingers was removed from the solar cell surface by selective wet etching. Photolithography was used for patterning in different fabrication steps. Some of the samples included an additional wet etching step, where approximately 25 nm of the contact GaAs was removed prior to the front contact fabrication. Conventional solar cells, i.e., having a linear contact grid pattern on top of the mesa structure without any insulating dielectric layers on the mesa facets were fabricated as references.

## 2.2 Characterization methodology

For studying the electrical performance of the solar cell components, light-biased IV (LIV) measurements were conducted using a 7 kW OAI TriSol solar simulator under AM1.5D spectrum (1000 W/m$^2$ normalization) at one-sun illumination and under 20–100 sun concentration. Current-voltage characteristics under dark conditions (i.e., dark IV) were measured to get more insight into loss mechanisms in the electrical operation. Two set-ups were used for the dark IV measurements: one with the OAI TriSol solar simulator and the other with a Keithley 2401 source meter unit. Both the LIV and dark IV measurements were conducted with 4-point probing.

The quality of the device processing was characterized by EL imaging, using a power supply and a microscope with a CCD sensor attached to a computer. The EL imaging was performed using different bias voltages and two different filters: a 750 nm short-pass filter (Thorlabs FES750) and an 800 nm long-pass filter (Thorlabs FEL800) for detecting EL from GaInP and GaAs subcells, respectively. The emission from the GaInNAsSb subcells (around 1300 nm) is beyond the detection range of the CCD sensor. In addition, the device components were imaged with a Carl Zeiss Ultra-55 scanning electron microscope (SEM).

Contact resistance measurements were conducted to investigate how the solar cell fabrication process affects the quality of the contact GaAs material and thereby also the quality of the metal-semiconductor interface. Transmission line measurement (TLM) method [15] was used for studying the effect of etching the contact GaAs layer at varying depths on the contact resistance. To mimic the actual solar cell fabrication process, the dielectric double-layer was first deposited and then etched completely with RIE from the whole front surface. The *n*-GaAs layer with a nominal doping level of $10^{19}$ cm$^3$ was etched to varying depths prior to the contact metal deposition by electron beam evaporation and the mesa fabrication. In addition, two reference samples were prepared: "Reference A" with no deposition nor etching of the dielectrics, and "Reference B" from which the dielectrics were etched with hydrofluoric acid instead of RIE to investigate the effect of PECVD on material quality. The values for contact resistivity were derived from measuring the IV characteristics of the contact pads having a size of 100 μm × 300 μm at varying distances according to TLM using Autolab PGSTAT101 Metrohm potentiostat.

## 3. Results and discussion

Table I summarizes the parameters derived from the one-sun LIV measurements. The values are averaged between the samples with different dielectric offsets. By comparing the values between the samples that had no preliminary contact GaAs etching (denoted as 'unetched') and the samples with the additional etching of contact GaAs prior to the front contact fabrication (denoted as 'etched'), a clear difference is seen. The fill factor and efficiency values are significantly higher for the samples with the additional etching step; the average fill factor values have increased from 77.3% to 83.8% and from 79.5% to as a high as 84.6% for SC1 and SC2 solar cells, respectively. This indicates that the additional contact GaAs etching prior to contact fabrication significantly improves the device performance, which can be observed also in the LIV characteristics shown in Fig. 2. Here, the shapes of the LIV curves between the 'unetched' and 'etched' SC1 components exhibit difference at higher voltages, where series resistance is important, indicating considerably higher series resistance for the unetched samples. In addition, the LIV graphs of the unetched SC1 components have an unusual shape, having a kink in the vicinity of the open-circuit voltage ($V_{OC}$). With the additional etching, this kink is completely removed, restoring good one-sun operation. This observation will be further discussed while analyzing the contact resistance results. Furthermore, similarly to SC1, pronounced series resistance effects are observed for SC2 in the unetched components compared to the components with the additional etching when considering the shape of the one-sun LIV graphs. In addition, with the unetched samples of SC2 a shunt-like behavior can be observed. However, this was associated with the wafer epitaxy and is thus not resulting from the fabrication process of the external busbars. As can be observed in Fig. 2, the reference component Ref-2 shows similar behavior in the area where shunt effects emerge while not demonstrating the pronounced series resistance effects.

*Table I. Average values of the parameters derived from one-sun LIV measurements shown together with standard deviations and number of samples (N). "Etched" refers to a process with an additional etching of contact GaAs prior to front contact fabrication, while "Unetched" refers to a process without any additional etching step.*

| Solar cell ID | | N | Efficiency (%) | $V_{OC}$ (V) | $J_{SC}$ (mA/cm²) | Fill factor (%) |
|---|---|---|---|---|---|---|
| **SC1** | Unetched | 3 | 24.4 ± 0.2 | 2.69 ± 0.01 | 11.7 ± 0.1 | 77.4 ± 0.7 |
| | Etched | 3 | 26.2 ± 0.2 | 2.67 ± 0.00 | 11.7 ± 0.0 | 83.8 ± 0.2 |
| | Ref-1 | 1 | 24.5 | 2.66 | 10.8 | 85.4 |
| | Unetched | 4 | 23.6 ± 0.6 | 2.63 ± 0.00 | 11.3 ± 0.1 | 79.5 ± 1.8 |
|     | Etched | 6 | 26.0 ± 0.5   | 2.64 ± 0.00 | 11.6 ± 0.2  | 84.6 ± 0.5  |
| SC2 | Ref-2  | 1 | 23.2         | 2.63        | 11.3        | 77.8        |
|     | Ref-3  | 1 | 24.6         | 2.64        | 10.8        | 86.2        |

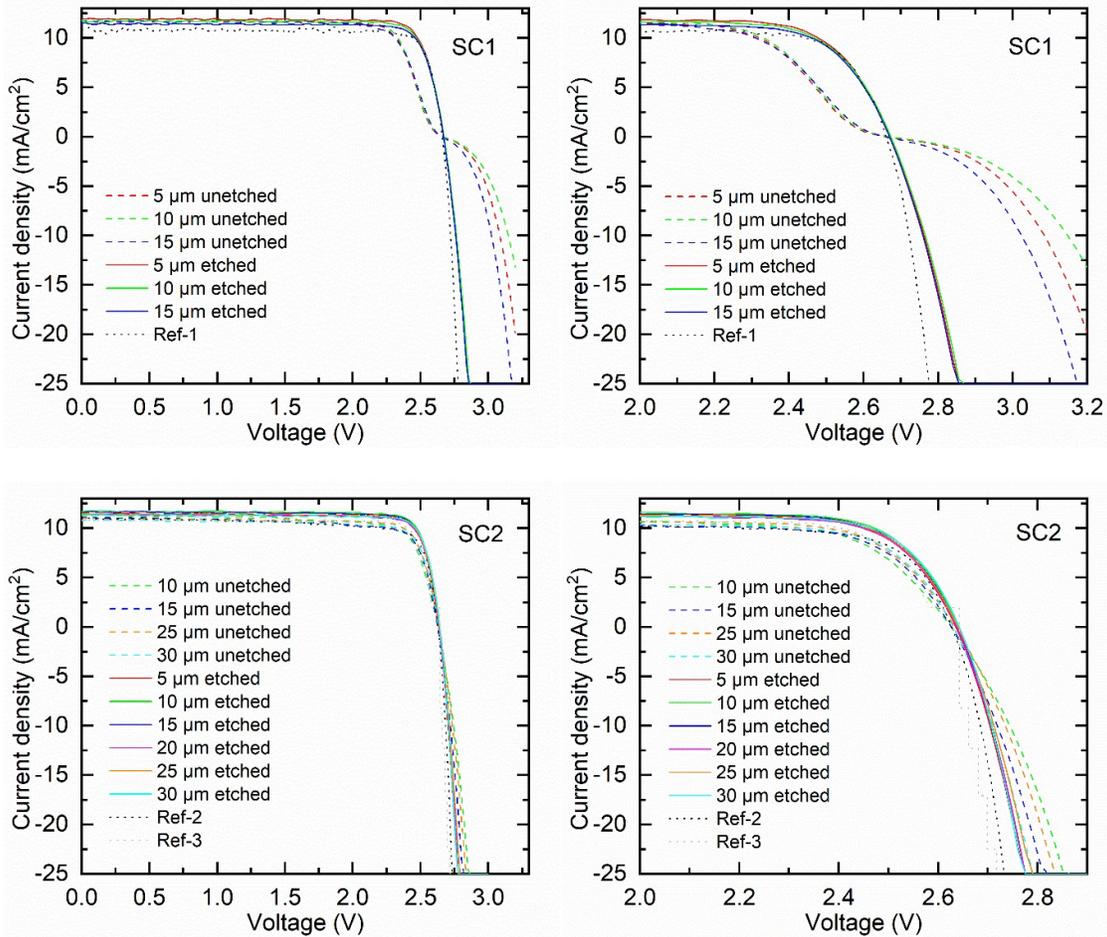

*Figure 2. LIV characteristics of the solar cells SC1 and SC2 with and without additional etching of contact GaAs (denoted as "etched" and "unetched", respectively) together with references.*

Overall, it can be concluded that the additional etching mitigates the issue with high series resistance encountered with the unetched solar cells. This can be further verified by examining the dark IV characteristics shown in Fig. 3. The area highlighted with a red circle in Fig. 5 is the region where the series resistance effects start being important for the cell operation. For both, SC1 and SC2, the highest series resistance is seen for the unetched components, but the series resistance is reduced for the etched components, whereas the reference components exhibit the lowest series resistance. It can be also seen in Fig. 3 that the difference between the unetched and the etched solar cells is more prominent for SC1 than for SC2, indicating that there is a larger difference in the series resistance between the etched and the unetched solar cells for SC1 compared to SC2. This is associated with SC1 exhibiting the unusual kink in the LIV graph.

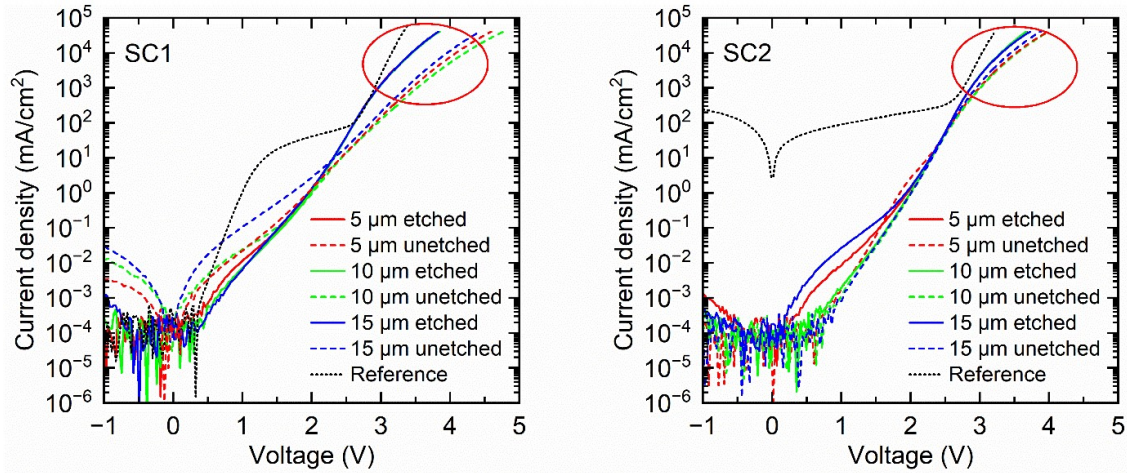

*Figure 3. Dark IV results of the unetched and etched solar cells SC1 and SC2 with dielectric offsets between 5 and 15 μm.*

The additional etching has negligible effect on the $V_{OC}$ and short-circuit current density ($J_{SC}$) values at one-sun illumination. Small differences may be accounted for normal variation between different device components resulting from fabrication. The higher average $V_{OC}$ of the unetched SC1 compared to the etched solar cells can be explained by the anomalous shape of the LIV graph. Regardless, the higher efficiency values resulting from the additional etching are mainly attributed to the higher fill factors and more ideal operation. Compared to the reference cells, the etched SC1 and SC2 components yield comparable performance, besides regarding the fill factor owing to the pronounced series resistance. Concerning the LIV performance under concentrated illumination, all the solar cell samples from SC1 and SC2, with or without the additional etching step, showed a kink in the LIV graph similar to the data shown in Fig. 4. This indicates that the emerging series resistance effects hinder the performance under high current conditions even when good one-sun performance was obtained with the additional etching step. When comparing the performance between SC1 and SC2 in Fig. 4, the characteristics related to series resistance seem to be stronger with SC1 as the kink is more pronounced. This is in line with the results discussed above, concluding that SC1 shows more pronounced series resistance effects.

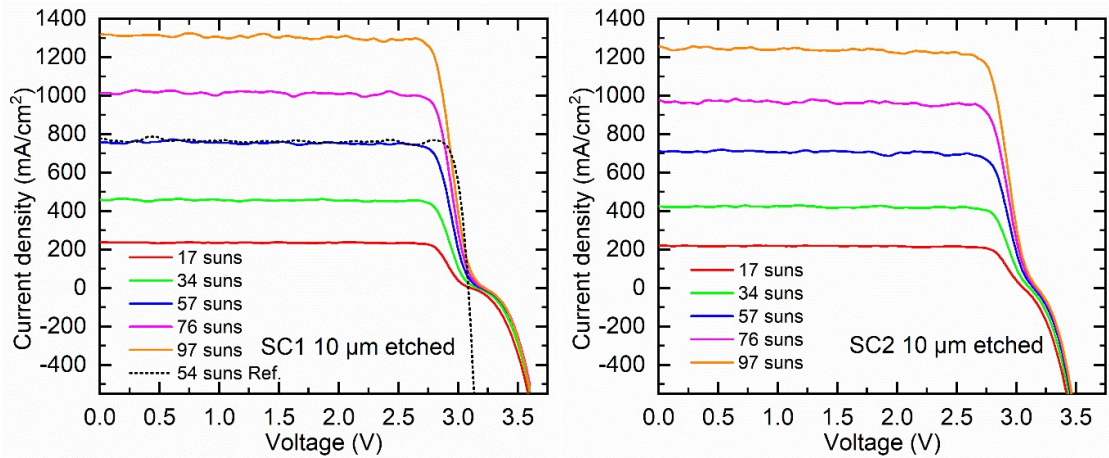

*Figure 4. LIV results under concentration for SC1 and SC2 with the dielectric offset of 10 μm and with the additional etching step.*

Analyzing the spatial distribution of EL shown in Fig. 5 supports the observations concerning the differences in the IV performance between the solar cells with and without the additional etching step. The EL from the GaInP subcell of the unetched solar cell is clearly nonuniform across the sample surface, whereas with the additional etching, significantly higher degree of uniformity in EL spatial distribution is detected from the GaInP subcell. In addition, to obtain similar current values in the EL measurement, higher bias voltages had to be applied for the unetched components compared to the etched ones, indicating additional resistive losses in the unetched components. These observations are associated with an uneven current distribution owing to pronounced parasitic resistive losses, also accounting for the decrease in the fill factor values. On the other hand, since the EL from the GaInP subcell of the etched samples is uniform throughout the sample, additional etching clearly remedies the issue that causes the uneven current distribution, suggesting that the issue lies in the top-most part of the contact GaAs layer. The EL originating from the GaAs subcell is uniform for both the unetched and etched solar cells, indicating that the issue causing the degradation in the electrical performance does not affect the GaAs junction.

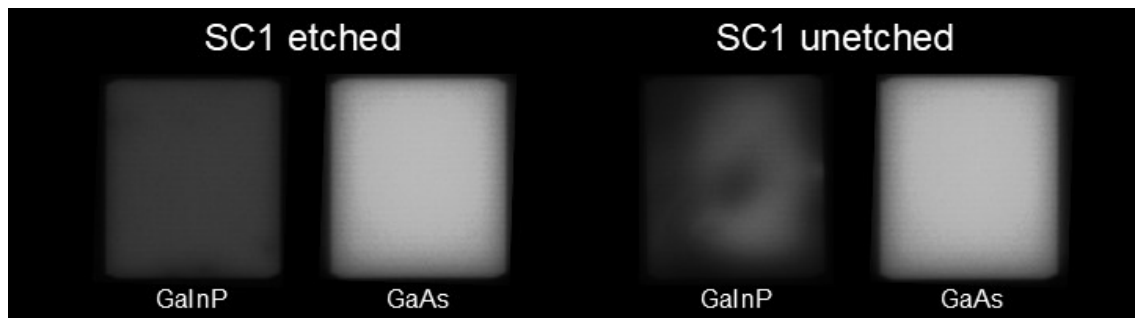

*Figure 5. EL images of the solar cell SC1 with and without an additional etching (denoted as "etched" and "unetched", respectively) with the dielectric offset of 5 μm.*

The IV measurements and EL characterization suggest that the issue causing the inferior electrical performance and pronounced series resistance effects is related to device processing steps, degrading the quality of the contact GaAs material. The "busbar-outside-mesa" process differs from a standard CPV process by the deposition and etching of the dielectric double-layer. Both of these steps are done with plasma tools, which are known to cause degradation of electronic properties in III–V materials [16–18]. To study whether the quality of contact GaAs was degraded during processing, a series of contact resistance samples with a varying etching depth of GaAs was prepared. As can be observed from the results presented in Table II, the obtained contact resistivity decreases when the GaAs etching depth increases. This gives a clear indication that the contact GaAs layer is nonuniform in depth. According to the results, etching as deep as 150 nm results in the contact resistivity values comparable to "Reference A", which had no deposition or etching of the dielectric double-layer, i.e., no interaction with RIE or PECVD. The type of the contact can be further characterized by the dark IV graphs shown in Fig. 6. The dark IV graphs of samples 1 and 2 are nonlinear, revealing a Schottky type behavior. The rest of the samples show linear IV behavior, indicating an ohmic contact formation. In addition, as the dark IV graphs of References A and B coincide, it can be stated that the damage originates from RIE etching, not from PECVD deposition. One explanation for this is that plasma damage originating from RIE alters the doping level of the contact GaAs layer, which alters the surface states in the material and thus also affects the formation of the contact in the metal-semiconductor interface. For samples 1 and 2 the damage is more extensive, as the formed contact is poorer forming even a Schottky barrier, which hinders the flow of the charge carriers across the metal-semiconductor junction. In addition, the electrical conductivity of the damaged $n$-GaAs is likely reduced. The damage depth in GaAs resulting from Ar plasma has previously been reported to reach even 120 nm in depth [16], which is well in agreement with our observations.

*Table II. Contact resistance results.*

| Sample ID | GaAs etch depth (nm) | Contact resistivity ($\Omega \cdot cm^2$) | Contact type |
|---|---|---|---|
| **Sample 1** | 0 | $2 \times 10^{-1}$ | Schottky |
| **Sample 2** | 25 | $1 \times 10^{-1}$ | Schottky |
| **Sample 3** | 100 | $4 \times 10^{-3}$ | Ohmic |
| **Sample 4** | 150 | $2 \times 10^{-4}$ | Ohmic |
| **Reference A** (no PECVD, no RIE) | 0 | $2 \times 10^{-4}$ | Ohmic |
| **Reference B** (PECVD, no RIE) | 0 | $2 \times 10^{-4}$ | Ohmic |

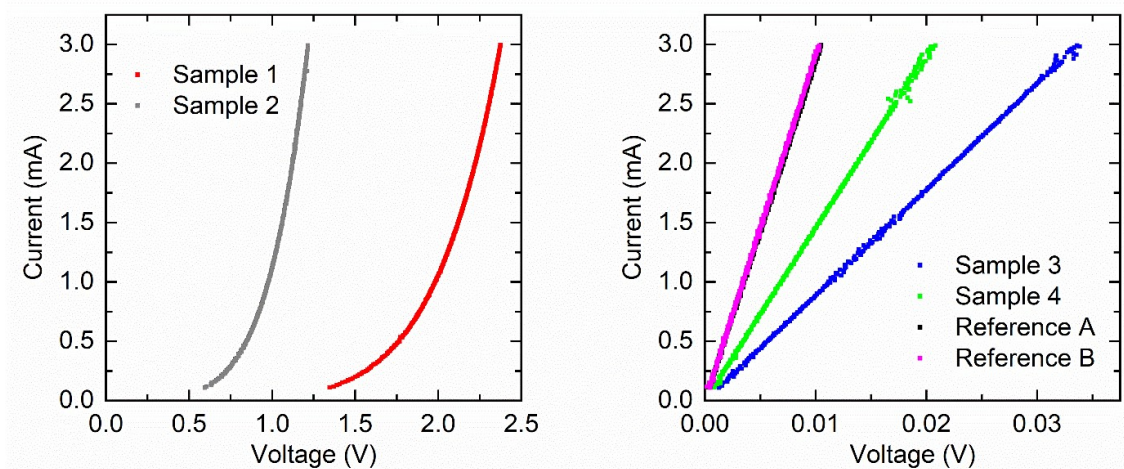

*Figure 6. Representative examples of the IV graphs obtained using TLM in contact resistance measurements to show nonlinear behavior for samples 1 and 2 (Left), and linear behavior for the rest of the samples (Right). Plots of References A and B are overlapping.*

The contact resistance samples 1 and 2 correspond to the case of the unetched solar cells and the solar cells with the additional etching step, respectively. This reveals the reason for the inferior one-sun LIV performance for the unetched solar cells: the contact resistivity is higher for the unetched solar cells compared to the ones with the additional etching step. Moreover, the kink in the one-sun LIV graph of the unetched SC1 in Fig. 2 is typically associated with a Schottky type behavior, which was proven to occur also with the contact resistance samples. Since the shape of the one-sun LIV graphs for the unetched SC2 is normal, i.e., it does not have the extra kink, it indicates that the quality of the electrical contact is better for the SC2 solar cells without the additional etching than for the SC1 solar cells. These observations are supported also by the dark IV measurements as discussed above. The reason for the difference may arise for example from the duration of RIE etching or some other aspect causing inconsistency in the fabrication process. Alternatively, contact GaAs doping may differ already as a result from the MBE growth, or the surfaces may have had dissimilar contamination before RIE process. Moreover, concerning the reference solar cells, those have not experienced the same plasma fabrication steps as SC1 and SC2 and should therefore correspond to the case of Reference A used for contact resistance measurements, having lower contact resistivity. This agrees with the reference cells showing lower series resistances in the LIV and dark IV measurements.

Overall, it can be concluded that etching approximately 25 nm of the contact GaAs is enough to ensure formation of an ohmic contact good enough for one-sun operation. However, for higher photocurrents seen in CPV measurements such etching remains insufficient, since the metal-semiconductor contact fails to operate due to resistive losses. To overcome the issue and to obtain properly functioning ohmic contacts, contact GaAs should be etched deeper. Also, to minimize material quality degradation, over-etching should be avoided once the dielectrics are completely removed from the surface. Furthermore, an alternative, more gentle dry etching recipe could be investigated. Wet etching would be a perfect choice for avoiding the material damage arising from plasma etching, but the isotropic etching profile with undercutting would be unfavorable in obtaining reproducible etching result and dimensional control. Another option to obtain ohmic contacts would be using annealing for the metals, but it would complicate the fabrication process as it requires two metallization steps, where the photolithography mask of the second step should align perfectly with the existing metal grid.

From the point of view of device fabrication, additional benefits are observed. Firstly, the process proved to be successful even with the smallest dielectric offset of 5 μm, which was not successful in the previous demonstration [13]. In an optimal case the dielectric offset should be fabricated as small as possible to avoid any additional power losses originating from its shadowing. This issue could be mitigated by using the most high-end mask aligner tools available to fabricate the photolithography for the dielectric opening. Another important observation concerning the device quality is that adhesion between metal and the dielectric double-layer was confirmed to be good since wire-bonding was successful on the busbars and no peel-off appeared.

## 4. Conclusions

The fabrication of an advanced front contact grid "busbar-outside-mesa" design was studied for GaInP/GaAs/GaInNAsSb triple-junction solar cells. It was revealed that etching of the contact GaAs layer prior to front contact deposition is required to obtain proper one-sun LIV performance and a uniform spatial distribution of EL. When etching approximately 25 nm of the contact GaAs, the average fill factor values increased considerably, namely from 77.3% to 83.8% and from 79.5% to 84.6% for SC1 and SC2 solar cells, respectively. In addition, there was a significant increase in the conversion efficiency values resulting from additional etching, increasing the values from 24.4% to 26.2% and from 23.6% to 26.0% for SC1 and SC2 solar cells, respectively. The LIV characteristics indicate that without the additional etching there are parasitic resistive losses, which was also revealed by the nonuniform distribution of EL in the GaInP subcells. This was concluded to result from material degradation, which was associated with the plasma damage originating from the fabrication of dielectric isolation double-layer. The damage reached as deep as over 100 nm but less than 150 nm according to contact resistivity measurements. In conclusion, to obtain ohmic contacts and proper device performance under concentration, the degraded material must be removed prior to contact fabrication. Alternatively, the damage could be avoided or at least diminished by avoiding over-etching or by using an alternative, non-degrading etching method. With further developments, the studied grid concept shows great promise to be employed in CPV and especially in micro-CPV, owing to the scalability and tunability of the design. The advantage of the design lies especially in minimizing the power losses originating from the grid shadowing and dark area related voltage losses.

**Acknowledgements**
The work has been funded by the European Research Council under the "AMETIST" ERC Advanced Grant ERC-2015-AdG 695116. The work is also part of the Academy of Finland Flagship Program PREIN 320168. The authors would like to thank Juuso Puutio and Seela Määttä for their technical assistance and Tampere Microscopy Center for SEM facilities. M.V.

acknowledges personal support from The Finnish Foundation for Technology Promotion, Jenny and Antti Wihuri Foundation, and Walter Ahlström Foundation.